\documentclass{cjaa}                    

\usepackage{graphicx}                   
\input{epsf.sty}                        
\input{psfig.sty}                       

\begin{document}

   \title{Understanding the  Chandra detected X-ray emission of the knots and hot spots of powerful  extragalactic jets.}



   \author{Markos Georganopoulos
      \inst{}\mailto{}
        }
   \offprints{}                   

   \institute{NASA/GSFC, Laboratory for High Energy Astrophysics, Greenbelt, MD 20771, USA
             \\
             \email{markos@milkyway.gsfc.nasa.gov}
                 }

   \date{Received ; accepted }

   \abstract{
I present here a short personal view of our understanding of the
Chandra detected knots and hot spots of powerful Fanaroff Rilley (FR) II
radio galaxies and quasars  in the context
 of  leptonic models. Observations of the  knots and hot spots
strongly suggest that the jets in these powerful sources retain their  
relativistic velocities at large scales, all the way to the hot spots.
The emission mechanism suggested  for the knots of quasars and FR II
radio galaxies is external Compton (EC) off the cosmic microwave backgounrd (CMB) 
from a relativistic flow,  while  for the hotspots Upstream Compton (UC) scattering from a 
decelerating relativistic flow.  
   \keywords{radiation mechanisms: non-thermal -- galaxies: quasars  }
   }

   \authorrunning{Markos  Georganopoulos}            
   \titlerunning{X-ray knots  and knots}  

   \maketitle
\section{Introduction}  

Jets, spanning distances up to $\sim$1 Mpc, emanate from the core of 
radio-loud active galaxies. Initially mapped at radio,
and recently, mostly through {\it HST} and {\it Chandra}, 
at optical and X-ray energies, they reveal a semi-continuous morphology
with bright knots and,  in the most powerful of them, hot spots, 
compact high  brightness regions, where the jet flow collides with the 
intergalactic medium (IGM).  Although the subject of intense studies
over the past decades, the physics of these objects is not well understood.
Little is known of  the mechanism of their origin,  or their 
composition (i.e. whether they consist of hadrons or leptons). 
Regarding the  kinematic state of the flows in
these objects (i.e. whether they are relativistic or not),
in both the  powerful FR  II/Quasars and the less powerfull
FR I sources the flow velocity of the pc-scale jets  is relativistic
with Lorentz factors $\Gamma \sim 10$. In the case of FR I sources
the flow seems to decelerate at kcp scales and the jet becomes
sub-relativistic (\cite{laing99}). 
Although there have long been   reasons to believe  (e.g. Wardle \& Aaron 1997) 
that the jets of the  powerful  FR II radio galaxies and quasars 
(which according to the unification scheme  are sources similar to FR II
radio galaxies but closer aligned to the line of sight; e.g. \cite{urry95})
remain relativistic at large distances from
the active nucleus, there has been no conclusive evidence for this to date.

Until recently, most of the extended jet observations were confined to radio 
interferometric studies, which probe the synchrotron radio emitting electrons
of the flow. In general it has been assumed that the radiating electron energy
density  is in   equipartition  with the magnetic field energy density,
because  equipartiton  minimizes the  energy content of the source
required for producing a given synchrotron power  and strong deviations from
equipartition can raise the energetic requirements to rather uncomfortable
levels.

Recently, astronomers were able to map
several extragalactic jets at optical and X-ray frequencies through HST and Chandra, with angular
resolution similar  to that of VLA.
In most, if not all cases where  optical emission was detected from a
radio knot or hot spot, the optical lies on  a  smooth 
or convex continuation  of the radio spectrum,  
suggesting that it is the extention of  the synchrotron spectrum. 
The situation with the X-ray flux of knots and hot spots 
 seems to depend on the source power. 
In general, in the knots of FR I sources 
the X-ray spectrum seems to be the high energy tail of the synchrotron
emission, altough, in a some  cases (e.g.  M 87; \cite{marshall02}),
 with a spectral break relative to the
radio-optical spectrum greater than the canonical value $\Delta \alpha =1/2$,
predicted my synchrotron cooling.
 In the more powerful sources, i.e. FR II radio
galaxies and quasars,   the X-ray
spectrum is not a continuation of the synchrotron spectrum,  
either because there is a cutoff
of the synchrotron spectrum at optical energies
 (e.g. in the hot spots of Cygnus A; \cite{wilson00}) and/or the X-ray flux lies
 above the extrapolation of the radio spectrum (e.g. in the knots of the quasar
PKS 0637-752; \cite{schwartz00}).
 The X-rays therefore must be the signature of another spectral component, 
possibly some kind of Compton scattering, given the presence of relativistic
 electrons in these environments.  
Two possible sources of seed photons for inverse Compton scattering 
are the synchrotron
photons produced in the source (synchrotron-self Compton, SSC) and external
photons, as  the photons of the CMB. In most cases, if  the 
emitting region is not moving relativistically and  the source is in 
equipartition, the SSC emission (SSC in equipartition, SSCE) dominates
over the EC emission.

Here I briefly review some of the Chandra knot and hot spot  observations  
of powerful sources (FR II radio galaxies and quasars) 
that critically affect our understanding of the powerful jet flows, 
and I show how, through studying two different locations, the knots and
 the hot spots, and using two
different lines of reasoning, the same conclusion is reached: {\it 
the large scale
jet flow remains relativistic up to $\sim$ Mpc scales, where the advancing jets
collide with the IGM.}
In \S\ \ref{equip}, I sketch  a simple derivation of the argument  
that the energy content of a  source emiting a given synchrotron luminosity
 is minimized when the 
electron energy density is in equipartition with the magnetic field energy
density.  I then use this result to show that the level of the SSC (EC) 
emission for a source in  equipartition decreases (increases) 
as the beaming of the source increases. 
In \S\, \ref{knots} I review the observations of the knots PKS 0637-752,
and the failure of the SSCE mechanism from a non beamed source   to
account for the bright X-ray emission of the knots (\cite{chartas00}). 
I then proceed to 
present the solution proposed independently by Tavecchio et al. (2000)
and  Celotti et al. (2001)':
the X-ray emission can be explained as EC scattering from a source in
equipartition if the source is moving with a substantial Lorentz
factor ($\Gamma\sim 10-15$) forming a small angle to the line of sight.
In \S\, \ref{hotspots}, following Georganopoulos \& Kazanas  (2003)[GK03] ,
after I review the disparate multiwavelength properties
of the hot spots of Cygnus A and Pictor A, 
I show that the two sources differ in 
orientation. I then show that all the sources with Chandra detected hot spots
can be broadly separated into Cygnus-like and Pictor-like and their properties
can be understood if the flow in the hot spots is relativistic and 
decelerating. This suggests that the flow feeding the hot spot is relativistic
and therefore the jets remain relativistic all the way to the hot spots. I 
also present a short discussion of UC scattering, a type
of scattering occuring in  decelerating flow, and I argue that it is
 responsible for the strong X-ray flux of the hot spots of Pictor A-like
 sources. Finally in \S  \ref{discussion}
I discuss some open issues and possible directions of future theoretical
and modeling work. 

\section{Equipartition and beaming}\label{equip}

{\bf Equipartition.} For simplicity, consider a synchrotron source with a 
monoenergetic electron population of a fixed  Lorentz factor $\gamma$.
The particle energy density is  $U_p= n\gamma m_e c^2$, 
where $n$ is the number density of electrons, $m_e$ is the electron mass
and $c$ is the speed of light. The energy density of the magnetic field $B$
is $U_B=B^2/8\pi$. The synchrotron luminosity is 
\begin{displaymath}
L_s\propto n\gamma^2 B^2=\gamma U_p U_B \nonumber
\end{displaymath}
For  fixed  $L_s$  the product $ U_p U_B$ is constant:
  $ U_p U_B=C$. The total energy density is  $U=U_p+U_B=U_p+C/U_p$
and it is minimized when 
\begin{displaymath}
{dU \over dU_p}=0 \Rightarrow U_p^2=C
\end{displaymath}
and since  $ U_p U_B=C$,
\begin{equation}
 U_p=U_B.
\end{equation}
Equipartition, therefore, between the radiating particles
and the magnetic field energy density minimizes the energy needed to 
produce a given synchrotron luminosity. 

{\bf SSC Luminosity As A Function Of Beaming.} The SSC luminosity
of  the source described above is
\begin{displaymath}
L_{SSC}\propto \gamma U_P U_s,
\end{displaymath}
where $U_s\propto\gamma U_pU_BR$ is the synchrotron photon energy density
of the source and $R$ is its fixed radius. The SSC luminosity can now be written
as 
 \begin{displaymath}
L_{SSC} \propto \gamma^2 R\, U_p^2 U_B 
\end{displaymath}
If now the source is moving with a bulk Lorentz factor $\Gamma$ forming
an angle $\theta$ to the line of sight, then the Doppler factor is 
$\delta=1/\Gamma(1-\beta\cos\theta)$ and the observed synchrotron luminosity
is $L_{s, obs}=\delta^4 L_s$. A  fixed observed synchrotron luminosity $L_{s, obs}$ implies a comoving synchrotron luminosity $L_s\propto \delta^{-4}$. Also, if the source is in equipartition
$L_s \propto U_B^2=U_p^2$, since $U_p=U_B$, and 
 $U_B=U_p\propto \delta^{-2}$.
Therefore $L_{SSC}\propto U_P^2U_B\propto\delta^{-6}$, and because $L_{SSC, obs}=L_{SSC}\delta^4$ one obtains
 \begin{equation}
L_{SSC, obs } \propto   \delta^{-2}.
\end{equation}
If a  source of a given
observed synchrotron luminosity  is assumed to be in equipartition 
the anticipated  SSCE luminosity decreases
by $\delta^{2}$ as the assumed beaming increases.  
This is the reason  relativistic
beaming was proposed as an explanation for the expected but not observed
high X-ray luminosities of compact radio sources.

{\bf EC Luminosity As A Function of Beaming.} If the source is propagating
through an isotopic photon fiend of energy density $U_{ext}$, its  energy
density in the source comoving frame will be $\Gamma^2U_{ext}$. 
The EC luminosity in the comoving frame will be 
$L_{EC}\propto \gamma U_p \Gamma^2 U_{ext}$. Given that 
$L_{EC,obs}=L_{EC}\delta^6/\Gamma^2$, and that in equipartition
 $U_p\propto \delta^{-2}$, the observed EC luminosity will scale with
beaming as 
 \begin{equation}
L_{EC, obs } \propto   \delta^{4}.
\end{equation}
 Therefore, for a source with a given
observed synchrotron luminosity, assumed to  be in equipartition, 
the observed EC luminosity increases by $\delta^{4}$ as the assumed 
beaming increases.

These results, demonstrated here for the limited case of a source with a 
monoenergetic population of electrons can be  appropriately generalized for a
power law electron energy distribution. The result that I will use in the 
rest of this review is that if the observed X-ray emission of a system is
higher that the estimated SSCE X-ray emission assuming no beaming, 
SSCE with beaming will reduce even further the  anticipated X-ray emission
relative to the unbeamed level. On the other hand, the anticipated 
EC X-ray level will increase with increasing beaming eventually reaching
the observed X-ray flux for an appropriately high value of $\delta$.

\section{PKS 0637-752 Knots: External Compton off the CMB}\label{knots}

{\bf SSC Is Out.}
One of the earliest  {\it Chandra} observations   was the detection 
of an one sided  large scale X-ray  jet in the 
quasar  PKS 0637-752 (\cite{schwartz00}, \cite{chartas00}), 
with strong emission form knots
located at  a projected distance of $\approx 100$ Kpc from the nucleus. 
The knots were also detected in the optical with {\it HST}, 
providing the opportunity to study their broadband radio to X-ray spectral
 energy distribution. The optical flux of the knots 
falls a factor of $\sim 10$ below the
line joining the radio and X-ray fluxes, indicating that there are two 
different spectral components and excluding the possibility that the entire 
spectrum is due to synchrotron emission. A natural candidate then for the
X-ray emission is SSC or EC scattering. However,  it was quickly pointed out
(Chartas et al. 2000) that,  in the absense of relativistic beaming,
the level of the X-ray  emission is  much stronger than what would be 
expected either from   SSCE ($\sim 2$ orders of magnitude) or 
EC scattering off the CMB photons ($\sim 4$ orders of magnitude)
assuming equipartition. Reproducing the X rays through SSC requires a magnetic
 field $\sim 50$ times below the equipartition level, which in turn increases
the energy requirements by $\sim  10^3$. An attempt to 
explain the X-ray flux through de-beamed SSCE also faces
severe problems: As shown in \S \ref{equip}
the SSCE flux increases with decreasing Doppler factor. Indeed,  
the X-ray flux of the knots can be reproduced assuming a Doppler 
factor $\delta=0.3$. This requires that the angle of the X-ray jet
 to the line of sight is $\approx 53^{\circ}$, given that explaining 
the X-ray non-detection of the counterjet requires a Lorentz factor 
$\Gamma \approx 8$
(\cite{schwartz00}). Such an angle however is in disagreement with the upper
limit $\theta<6.4^{\circ}$ of the VLBI jet derived from the observed 
superluminal motions   (\cite{lovell00}), and, given that there is no apparent
bend between the VLBI and large scale jet, the only permitted configuration
would require a bend jet on a plane perpendicular to the plane of the sky.
As Schwartz et al. (2000) pointed out, even in this remote case the apparent
 radio
luminosity of $3 \times 10^{43}$ erg s$^{-1}$ would correspond to a radio
luminosity of  $2 \times 10^{46}$ erg s$^{-1}$ for a similar source pointing
 toward the 
observer, exceeding significantly the blazar radio luminosity. 
It seems therefore
that SSC emission is not a viable mechanism for the observed X-ray emission
of this source.

{\bf EC Is In.} The next natural candidate for reproducing the X-ray flux
is EC scattering off  CMB photons. In this case, as was demonstrated
in \S \ref{equip}, an increase of the Doppler factor increases the level
of the EC emission. EC scattering is also promising  because the jet of 
this particular source forms a small angle to the line of sight, resulting
to a substantial value of $\delta$ for a given $\Gamma$. 
Tavecchio et al. (2000) and Celotti et al. (2001) showed that if
the knot plasma is flowing relativistically with a Lorentz factor 
$\Gamma \sim 10$, forming an angle $\theta\approx 5^{\circ}$ to the line
of sight, the boosting of the CMB photon energy density by $\Gamma^2$ in the
comoving with the flow frame, reproduces the observed X-rays through EC scattering, with the source being in  equipartition.
According to this picture, the electron energy distribution must have a low
energy cutoff at $\gamma\approx 10$, otherwise it would overproduce the optical
flux (Tavecchio et al. 2000). The total jet power required in this picture
is of the order $10^{47-48}$ erg s$^{-1}$, similar to the power carried by the
jets of the most powerful blazars. A comparison of the synchrotron to the 
model EC luminosity shows that in this scenario the EC losses dominate the
energetics. At larger angles however, as $\delta$ decreases, the EC flux
decreases and the SSC flux increases relative to the synchrotron flux assuming equipartition. Therefore, similar sources at larger
angles, will have an X-ray output dominated by SSC emission, although the
radiative losses will still be dominated by EC scattering. This is an
interesting point that has to be taken into account in modeling such sources
at large angles: one has to include the EC losses in the calculation of
the energy losses and of the electron enectron energy distribution,
even though SSC emission  may dominate over EC emission at large angles.

{\bf For Ever Detectable.}
An interesting corollary of this picture was recently described by
Schwartz (2002): if the X-rays are due to EC scattering on the CMB,
then these knots should be  detectable with the same surface brightness
even if the source is located at much higher redshifts $z$. 
This is because the CMB
energy density scales as $(1+z)^4$ and exactly compensates the $(1+z)^{-4}$
scaling of the surface brightness 
(note that the angular size remains
practically constant for $z>1$, almost independently of the cosmology chosen).
Therefore, if the X-ray knots are due to EC scattering from the CMB,
they should be observable at any redshift at which they exist. 
On the other hand, the flux from the core, assumed not to depend on the
 CMB energy density, will decrease with increasing luminosity distance 
(redshift), and the X-ray observed flux will be dominated by the knots  and not
by the core at redshifts above $z\approx 3-4$, contrary to  the case of
 nearby sources like PKS 0637-752 and 3C 273 (\cite{marshall01}).
Then, at large $z$, one expects to see in X-rays only  the jet - knot emission,
 displaced by the optical core (non-detectable in  X-rays) by $5^{\arcsec}-10^{\arcsec}$. These 
X-ray bright-radio quiet jets should  be among the unidentified ROSAT 
sources, and their detection or not  will provide a test for the EC knot emission model.

{\bf The Cooling Problem: Continuous Jets, Not Knots.} 
Although the radiative cooling length of the electrons responsible for the
synchrotron optical and EC $\gamma$-rays is $\sim 10$ Kpc (\cite{tavecchio00}),
comparable to the $\sim 3 $ Kpc size of the PKS 0637-752 knot size, as
Schwartz (2002) noted,  the radiative cooling length of the low energy
electrons ($\gamma\sim 100$) that produce the X-rays through EC
scattering off the CMB is $\sim 100$ Kpc or more, comparable to the 
total length of the jets, and much larger than the typical observed knot sizes.
Therefore, instead of the observed knot morphology,  according to the EC 
scenario one should observe a continuous jet in X-rays.
Tavecchio at al. (2003) note that  outside the bright knots 
X-rays dim as fast as  optical and radio, and that this behavior
is not compatible with radiative losses, which would give at each
frequency a size proportional to the corresponding electron loss length scale.
They then suggest  adiabatic losses as the dominant electron energy loss
mechanism for all but the most energetic electrons. The attractive 
characteristic of adiabatic energy losses  is that they cool also the low energy 
electrons responsible for the EC X-ray emission, thus reducing  the
size of the X-ray emitting regions. However, for typical values of the
physical parameters involved they find that the size of the X-ray emitting
regions is still much larger than  observed.  
To overcome
this problem and keep the EC/CMB interpretation of the X-rays, they
speculated that each observed knot is a collection of  micro-knots which expand
adiabatically in three dimensions, resulting  to strong adiabatic 
electron losses that keep the cooling length of the  X-ray emitting 
electrons roughly equal to the observed knot size. 

A plausible solution to the problem of the size of the knots
(Georganopoulos \& Kazanas, in preparation) 
is that low energy electrons are present throughout 
the jet and   there  is no need for them to cool  catastrophically. 
The reason then 
we observe a knot morphology in X-rays   is that the needed  seed photon
energy density is present only in the knot environment. If the flow in the 
knots is relativistic and decelerating the electrons of the faster part of
the flow will see the synchrotron emission of the slower part of the knot
relativistically boosted and will radiate  X-ray emission 
(more on this type of Compton  emission, in \S \ref{hotspots}). 
Once they are advected away from the fast part of the flow, they stop 
radiating in X-rays, not because they do not have the necessary energy,
but because they lack the seed photons for Compton scattering.
In addition EC off the CMB will be more effective and beamed  at the fast part of the flow,
where the Lorentz factor is higher, decreasing its power and widening its beaming pattern
at the downstream part of the flow. 
Given that the knots are sites of particle deceleration, which most probably
takes place in shocks, it is natural to assume that these  shocks are the
locations of the needed bulk flow deceleration.

\section{Hot spots}\label{hotspots}

The first hot spots to be detected in 
X-rays were those of the nearby powerful radio galaxy Cygnus A 
(Harris et al. 1994), whose X-ray flux measured by {\it ROSAT} 
was found to be in agreement with 
SSCE.
Because the advance
speed of the hot spots through the IGM is slow 
($u/c\approx 0.1$, e.g. Arshakian \& Longair 2001),
 it has been implicitly assumed that the plasma flow
in the hot spots is also sub-relativistic.
 The Cygnus A hot spots  show no optical emission, 
suggesting a synchrotron spectral cutoff at lower frequencies. 
This was therefore the situation before the Chandra era, as confirmed
by {\it ROSAT} observations of Cygnus A: 
the plasma in the hot spots  was assumed to be in equipartition 
and moving sub-relativistically. 

While {\it Chandra} observations of Cygnus A confirmed the SSCE picture 
(Wilson et al. 2000), observations of Pictor A, another nearby powerful 
radio galaxy, present a drastically different picture  (Wilson et al 2001): 
($i$) An one-sided large scale X-ray jet on the same direction 
with the known VLBI jet (Tingay et al. 2000). 
($ii$) Detection of X-ray and optical
emission only from the hot spot on the jet side. ($iii$) SSCE models for the Pictor A 
hot spots under-produce the observed X-ray flux: The  
synchrotron photons are  not a sufficient source of seed photons for
producing the observed SSC  emission under equipartition
conditions.  This last point describes 
the {\it problem of the missing seed photons}, which 
 is likely to be germane to other high energy sources (e.g. TeV Blazars; Georganopoulos \& Kazanas 2003b):
{\it The synchrotron photon energy density in the SSCE model is lower
that the seed photon energy density required  to produce the observed
inverse Compton emission.}  
In the SSC model a decrease in the magnetic field leads to an 
increase of the  Compton to synchrotron flux ratio. In the case of Pictor A,
a magnetic 
field $\sim 14$ times below its the equipartition value is required 
in order to 
achieve agreement with the observed the X-ray flux. 
The problem of 
the disparate  hot spot properties of Cygnus A and Pictor A deepened 
further with the discovery by {\it Chandra} of more X-ray emitting 
hot spots, and the issue is currently a matter of active debate (e.g. Wilson 2001, Hardcastle et al. 2002, Kataoka et al. 2003, GK03).

{\bf An Orientation Sequence.} 
An indicator of orientation for radio-loud active galaxies
 is the ratio $R$ of the core (beamed)   to the extended (isotropic)
  radio emission.
 Sources with jets closer to the observer's line of sight are expected to
 have higher values of 
$R$ than those with jets on the plane of the sky. Cygnus A has 
$\log R \approx -3.3$, while Pictor 
A has  $\log R\approx -1.2 $, suggesting that the jets 
of Pictor A are closer to the line of sight than those of Cygnus A. 
Another indicator of source orientation is the detection of broad
emission lines in the optical-UV spectrum of the core emission. 
According to the unification scheme for radio  loud active galaxies
(e.g. Urry \& Padovani 1995),  broad line radio galaxies (BLRG) and  quasars 
have jets pointing close to the line of sight, while narrow line 
radio galaxies (NLRG) have jets closer to the plane of the sky. 
Cygnus A is a NLRG, and Pictor A is a BLRG, suggesting again that 
Pictor A is aligned  closer  to the line of sight.  

Motivated by the orientation difference of the  two sources,  
GK03  studied the sources with X-ray hot spot
 detections,
mostly from  high resolution {\it Chandra} observations.
A set of correlations with jet orientation emerges 
from this study: \\
$\bullet$ As the jet alignment with the observer's 
line of sight manifest by $R$ increases, the source changes from a NLRG 
to a BLRG/Quasar. \\
 $\bullet$ Sources with jets closer to the plane of the 
sky show hot spots of comparable X-ray flux from both lobes, 
while sources with more aligned jets show X-ray 
hot spots on the side of the near jet, as identified through VLBI 
observations. \\
  $\bullet$ While in NLRG the radio-to-X-ray spectra are 
modeled successfully by SSCE, in the more aligned BLRG and quasars, 
the hot spot X-ray emission is significantly stronger than its SSCE predicted 
value, again a manifestation of the {\it problem of the missing seed photons}
 facing SSCE models;  Hence, in the SSC framework 
one has to resort to magnetic fields well
 below equipartition by factors $\sim 10-30$
 to reproduce the observed X-ray flux, 
increasing by orders of magnitude the required jet power. \\
   $\bullet$ 
Synchrotron optical emission,  weak or absent in the NLRG hot spots, 
appears at the jet side hot spot as the source aligns closer to the 
line of sight. \\
 $\bullet$ Radio emission is seen  from both 
hot spots, regardless of orientation, although the hot spots on the VLBI 
jet side of BLRG/quasars are brighter and have a flatter spectrum.

{\bf Differential Beaming.} 
GK03 proposed
that this  orientation sequence  can be accounted for by appealing to 
{\sl frequency dependent} 
beaming of the hot spot emission. 
In this case, the synchrotron emission is beamed most
at its highest (optical) frequencies. As a result it is observed
preferentially in the near hot spot of the more aligned objects 
(BLRG/quasars). Beaming decreases (i.e. the intensity amplification becomes
lower and the beaming pattern broader) with decreasing frequency, with the 
lowest frequency (radio) emission being the least beamed, thus observed 
in the hot spots of both jets of all objects (albeit with 
higher flux from the near jet hot spot).

The X-ray emission is generally attributed to the Compton component of 
the SSC process by the electrons responsible for the radio emission. 
As such,  the orientation dependence of these two components should be 
virtually identical. This is in agreement with the X-ray detection
from both hot spots in NLRG and their successful modeling with 
SSCE. It would then appear that the X-ray detection from single
hot spots, as the source alignment increases, is at odds with 
this picture. However,  detections of 
X-rays from only the near hot spot are  also accompanied by an
 increase in the radio-to-X-ray ratio beyond that of the 
SSCE models, indicating the presence of an additional component 
more sensitive to orientation  than SSCE. 

{\bf Decelerating Relativistic Flows and Upstream Compton Scattering.}
GK03 argued  that  
this frequency dependent beaming and the excess X-ray emission in the 
more aligned sources can be accounted for in terms of a relativistic, 
decelerating flow at the hot spots. This possibility has not been 
explored in view of their sub-relativistic ($u/c\approx 0.1$, e.g. Arshakian \& Longair 2000) advance speed through the IGM. However 
hot spot flow patterns with Lorentz factors  up to $\Gamma\sim 3$, 
decelerating to the sub-relativistic hot spot advance speed 
are routinely seen  in relativistic hydrodynamic simulations 
(Aloy et al. 1999, Komissarov \& Falle 1996), 
and have been invoked (Komissarov \& Falle 1996) 
to account for the observed higher radio brightness of emission from 
the near hot spots compared to that of the far ones.

 In such decelerating flows,
the highest frequencies of the  synchrotron component originate 
at thw fast base of the flow where the 
electrons are more energetic and its Lorentz factor largest. As both 
the flow velocity and electron  energy drop with distance, the locally emitted 
synchrotron spectrum shifts to lower energies while its beaming pattern 
becomes wider. The observed synchrotron spectrum is the convolution of the 
comoving emission from each radius weighted by the beaming amplification 
at each radius, with the high energy photons 
beamed in a more narrow angular pattern than the low energy   ones.

The inverse Compton emission of such flows behaves in a more involved
way: Electrons  upscatter the locally produced synchrotron seed photons,
giving rise to a local SSC emission with a beaming pattern  identical to 
that of synchrotron.  In addition to this, Upstream Compton (UC) takes place,
a process in which synchrotron photons from its slower
downstream section  serve as  seed  photons for Compton  scattering  by 
electrons  in its upstream faster part.  
 The energy density of these downstream produced synchrotron photons, 
is   boosted in the fast (upstream) part of the flow by $\sim 
\Gamma_{rel}^2$ (Dermer 1995), where $\Gamma_{rel}$ is the relative 
Lorentz factor between the fast and slow part of the flow.
Also, the  beaming pattern of the UC 
radiation is narrower  (see Appendix   \ref{UCbeam})
 than that of the  synchrotron/SSC pattern
of $\delta^{2+\alpha}$, where $\alpha$ is the radiation spectral index, 
approaching the   $\delta^{3+2\alpha}$  beaming pattern
of external Compton (EC) scattering  
 (Dermer 1995, Georganopoulos et al. 2001).
It is the combination of higher synchrotron photon energy density
measured at  the comoving with the fast flow frame and the more narrow
than synchrotron/SSC beaming of UC scattering that explain why the
additional X-ray emission, which gives rise to the missing
seed photon problem in the framework of SSCE models, 
is evident only in the more aligned sources.

The following simplified model, involving relativistic 
decelerating flows, was used to reproduce the multi-object, multi-frequency 
phenomenology of hot spot emission: A relativistic EED
$N_e(\gamma) \propto \gamma^{-s}, \, s=2$ with high energy cut-off 
at $\gamma_{max} \simeq 2 \cdot 10^6$ is injected impulsively at 
the base  the flow, in equipartition with the ambient magnetic field 
and assumed to remain so as the flow decelerates. A plane-parallel 
flow geometry  is assumed with bulk Lorentz factor decreasing 
with the distance $z$ from the shock as $\Gamma(z) = \Gamma_0 
(z/z_0)^{-3}$, with $\Gamma_0=3$ and $z_0 \simeq 1$ kpc, in agreement 
with {\it Chandra} observations. 
The synchrotron and Compton emission coefficients are computed using
approximate formulae  and the 
hot spot spectra are obtained by integrating these over the volume of 
the flow taking into account both their $z-$ and $\theta-$ dependent 
beaming amplification ($\theta$ is the angle between the flow velocity 
and the observer's line of sight) assuming a ``pill-box" geometry for the 
flow, i.e. transverse dimension $D = 2 z_0$. 
The resulting spectra 
under typical BLRG/Quasar  ($\theta = 20^\circ$,  solid line) and NLRG ($\theta = 70^\circ$, dashed line) orientations are given in figure 1.

\begin{figure}
   \vspace{2mm}
   \begin{center}
   \hspace{3mm}\psfig{figure=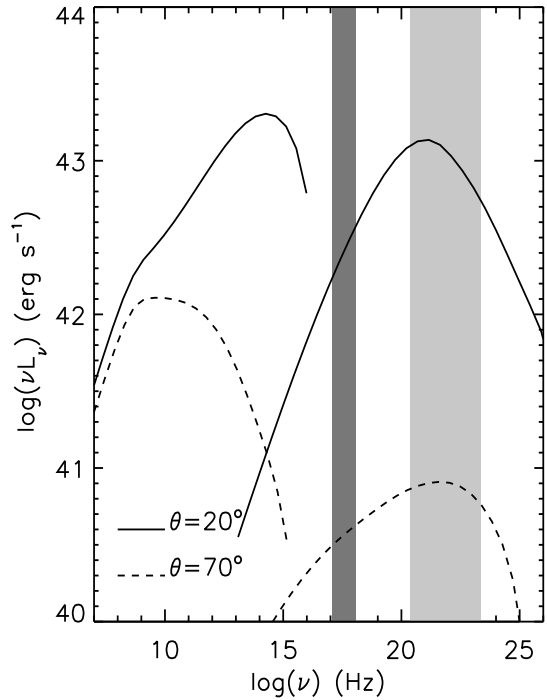,width=80mm,height=110mm,angle=0.0}
   \caption{ The synchrotron and inverse Compton emission of a decelerating relativistic flow observed under two different angles.}
   \label{Fig1}
   \end{center}
\end{figure}

Several important points  are 
apparent in this figure: \\
$\bullet$ The ($\theta = 70^\circ$) spectrum 
exhibit an X-ray to radio ratio very similar to 
that observed in Cygnus A and the rest of the NLRGs, 
while, the ($\theta = 20^\circ$) spectrum, has  a significantly higher
X-ray to radio ratio due to UC scattering, in agreement with those
observed in BLRG/quasars, including Pictor  A.\\
 $\bullet$ The 
synchrotron component is prominent at optical frequencies for
$\theta = 20^\circ$, while essentially cuts off for $\theta = 
70^\circ$, in agreement 
optical hot spot detections only from the hot spots of the approaching jet.\\
$\bullet$ The change of slope at the ``cooling" break at $\nu \simeq 1$ 
GHz is much smaller for $\theta = 20^\circ$ and generally 
consistent with the spectrum observed in Pictor A (Wilson et al. 2001),
while at  $\theta = 70^\circ$ there is a stronger break  similar to that 
observed in Cygnus A, and with a slope  with the canonical 
value $\Delta \alpha = 1/2$. \\
 $\bullet$The synchrotron spectra of the hot spots
viewed at smaller angles are {\sl flatter} than those at large
angles, and also flatter than those of the receding jets (not shown here).
This is a well known fact (Dennett-Thorpe et al. 1997)
 but without any clear explanation
todate. Decelerating flows provide a straightforward account
for it: At small angles,  the decelerating 
flow  amplifies the emission of higher synchrotron frequencies more than 
that of lower frequencies, leading
to flatter spectra than those observed at large angles for which
relativistic boosting is not significant. For the same reason,
  the break at large angles and/or rapid decelerations
can increase  up to $\Delta \alpha  \approx 0.7$, something that has been
observed in the X-ray detected knots of the FR I galaxy M87 
(Marshall et al. 2002).\\
 $\bullet$ 
In some of the nearby ($z\leq 0.1$) 
Pictor A- type sources, the hot spot high energy $\gamma-$ray 
flux  in the 10 MeV - 10 GeV band of {\it GLAST} will be both detectable
 and its position sufficiently well determined to be distinguished 
 from the (potential) emission  from the AGN ``core''. \\

I stress that the existence of relativistic flows in the hot spot 
requires that the jet flow remains relativistic up to its termination  at the hot spots at distances up to $\sim $ Mpc  from the AGN ``core''.

\section{Conclusions}\label{discussion}

{\it Chandra} offers us the unique opportunity to study extragalactic jets
in X-rays with arcsecond resolution. This capability resulted to the discovery
of an increasing number of X-ray emitting knots and hot spots in extragalactic
jets. The study of both the knots and the hot spots of powerful jets support 
the idea that these jets remain relativistic all the way to the hot spots,
where the jets collide with the IGM. Previous arguments for large scale
relativistic flows (e.g. Wardle \&  Aaron  1997) were based on radio data only.
It now seems safe to state that   sub-relativistic large scale flow velocities  
in the  jets cannot reproduce the multi-wavelength multi-object 
observations of  knots and hot spots.

Assuming that, aside the idiosyncracies of individual sources, all powerful extragalactic
jets are at first order similar, a unified phenomenological framework, based on the orientation of the sources,
seems promising; such a scheme should in the future also include the source power to accomodate the knots
of the less powerful sources like FR Is and their aligned population BL Lacertae objects (in these sources 
the knot X-ray emission seems to be dominated by synchrotron radiation).
A similar situation emerges in blazars, where a recent study by Padovani et al. (2003) shows that powerful
blazars  avoid the extreme synchrotron peak frequencies that  BL Lacertae objects can reach.  

From the theoretical/modeling point of view the first step to be taken is to establish the energy loss and
radiation mechanisms responsible for the broadband spectra of hot spots and knots. Since the source 
orientation is involved, it is imperative to have observations of several sources with jets oriented at different angles,
and, if possible, observations of the counter jet/counter hot spot. Note here that, while inverse Compton scattering
either as  EC off the CMB or UC  may be the dominant energy loss mechanism, if the source is misaligned the X-ray output
will mostly be due to SSC. The broadband hot spot spectra   of broad line (i.e. closer aligned) objects as
well as narrow line (i.e. closer to the plane of the sky) objects, seem to fit in the scheme of GK03, according to which the flow in the hot spots is relativistic and decelerating.  In this scheme the higher than anticipated  X-ray flux
of the broad line objects is attributed to UC scattering. Further {\it Chandra} observations will check and quantitiatively constrain this
scheme. 

The knot broadband emission of the aligned superluminal quasar PKS 0637-752 has been interpreted by Tavecchio et al.
(2000) and Celotti et al. (2001) as EC scattering off the CMB background. A serious difficulty this scheme  encounters
is the fact that the cooling length of the $\gamma \sim 100$ X-ray emitting electrons is much larger that the observed
knot size. A solution of this problem can be reached assuming either that the electrons cool catastrophically inside the knot, or that the electrons just lack the appropriate seed photons outside the knot. The first alternative prompted
Tavecchio et al. (2003) to suggest that each knot is composed of several micro-knots and that adiabatic losses cool the electrons of these micro-knots below  $\gamma \sim 100$, after they have propagated for a distance similar to the macroscopic hot spot. The second alternative naturally occurs in a decelerating flow,
where both the UC and EC emission off the CMB  are confined at the fast base of the flow (EC emission will obviously
be produced downstream, albeit at a lower total power, since the comoving CMB photon density scales as $\sim \Gamma^2$,
and with a wider beaming pattern. These alternatives give different variability and orientation behaviors which can be tested through future observations.

After establishing the radiation mechanisms, an interesting question, similar in nature to the one still being open in
blazar research should be addressed: why the synchrotron spectrum of the most powerful sources (cores, knots, hot spots)
cannot reach X-rays energies? In the case of the cores of the powerful blazars , it was suspected that this is due to the photon field of the  broad line region increasing the radiative losses of the electrons and reducing the maximum electron energy. However this cannot be the explanation for the extended jet features that are far from the broad line photon fields. This opens the possibility that the spectral difference between weak and powerful sources
reflects an intrinsic different that persists throughout the extended jet. 

The ground for studying  the large scale extended jets is very fertile. Our new observational capabilities offer a foundation  we can use to advance our understanding of extended jets. Hopefully, this will bring within our reach
the solution of an old persistent problem, that of the matter content of the jets.

\begin{acknowledgements}
 I want to thank my NRC adviser Demosthenes Kazanas for a fruitful and stimulating collaboration
and the organizers of the Vulcano meeting for the invitation and for organizing an excellent meeting.

\end{acknowledgements}
\appendix                  

\section{The beaming of UC scattering}\label{UCbeam}
I outline here the argument that the UC beaming is more
tight than the Synchrotron/SSC beaming.
Consider for simplicity
a two-zone flow, a fast part with Lorentz factor $\Gamma_1$ followed by a
 slower
part with Lorentz factor $\Gamma_2$. Consider also an observer
located at an angle $\theta$ such that  the  Doppler factors of the
two zones are $\delta_1 $, $ \delta_2$.  The beaming pattern of the UC 
radiation in the frame of the slow part of the flow will be 
$\delta_{1,2}^{3+2\alpha}$, where 
$\delta_{1,2}$ is the Doppler factor of the fast flow in the frame of the
slow flow. To convert this beaming pattern to the observer's frame
we need to boost it by $\delta_2^{2+\alpha}$. The beaming pattern is then
written as $\delta_{1,2}^{3+2\alpha}\,\delta_2^{2+\alpha}$. To write
$\delta_{1,2}$ as a function of $\delta_{1}$, $\delta_{2}$, we note that
a photon emitted in the fast part of the flow is seen by the observed
boosted in energy by a factor $\delta_1$. The same boosting can take place
is two stages: first going to the frame of the slow flow by being boosted 
by  $\delta_{1,2}$ and then going to the observer's frame by  being boosted 
by $\delta_2$. Because  the final photon energy in the observer's frame does
not depend on the intermediate transformations,  $\delta_{1,2}=\delta_{1}
/\delta_{2}$.  The beaming pattern of UC scattering is therefore 
$\delta_1^{3+2\alpha}/\delta_2^{1+\alpha}$.   Note that, as expected, 
 for $\delta_1=\delta_2$, we recover the beaming pattern of SSC, 
while for $\delta_2=1$, that of EC radiation.

\label{lastpage}

\end{document}